\documentclass[sigconf, nonacm]{acmart}

\theoremstyle{definition}
\newtheorem{definition}{Definition}
\newcommand{\knn}{\textsc{Knn}}
\newcommand{\q}{\mathfrak{q}}
\newcommand{\Q}{\mathcal{Q}}
\newcommand{\I}{\mathcal{I}}

\newcommand{\memory}{$60\%$}
\newcommand{\perf}{$30\%$}
\newcommand{\recall}{$2\%$}

\usepackage{booktabs}   
\usepackage{subcaption} 

\begin{document}

\title[]{Low-Precision Quantization for \\ Efficient Nearest Neighbor Search}         


\author{Anthony Ko}
\affiliation{%
  \institution{Amazon}
  \country{}
}
\email{anthoko@amazon.com}
\author{Iman Keivanloo}
\affiliation{%
  \institution{Amazon}
  \country{}
}
\email{imankei@amazon.com}

\author{Vihan Lakshman}
\affiliation{%
  \institution{Amazon}
  \country{}
}
\email{vihan@amazon.com}
\author{Eric Schkufza}
\affiliation{%
  \institution{Amazon}
  \country{}
}
\email{eschkufz@amazon.com}

\begin{abstract}
Fast $k$-Nearest Neighbor search over real-valued vector spaces (\knn{}) is an
important algorithmic task for information retrieval and recommendation systems.
We present a method for using reduced precision to represent vectors
through quantized integer values, enabling both a reduction in the memory
overhead of indexing these vectors and faster distance computations at query time. While most traditional quantization
techniques focus on minimizing the reconstruction error between a point and its
uncompressed counterpart, we focus instead on preserving the behavior of 
the underlying distance metric. Furthermore, our quantization approach is applied at the implementation level and can be combined with existing \knn{} algorithms. Our experiments on both open source and proprietary datasets across multiple popular \knn{} frameworks validate
that quantized distance metrics can reduce memory by \memory{} and improve query
throughput by \perf{}, while incurring only a \recall{} reduction in
recall.

\end{abstract}


\maketitle
\pagestyle{plain}
\section{Introduction}
\label{sec:introduction}

\knn{} search has become increasingly prevalent in machine learning applications and
large-scale data systems for information retrieval~\cite{huang2013learning} and
recommendation systems targeting images~\cite{10.1145/1862344.1862360}, audio~\cite{7894058},
video~\cite{6642782}, and textual data~\cite{10.1007/978-3-642-15286-3_16}. The classical form of the problem
is the following: given a query vector $\q \in \mathbb{R}^{d}$, a distance metric $\phi$, 
and a set of vectors $\I,$ where each $x \in \I$ is also in $\mathbb{R}^{d},$ find the set of $k$
vectors in $\I$ with the smallest distance to $\q$. 

Most methods for improving the performance of \knn{} focus on pruning the
search space through the use of efficient
data structures~\cite{teflioudi2016exact:lemp,li2017fexipro,johnson2019billion:faiss,garcia2008fast:knn-gpu}.
Some methods also use compressed vector representations that trade a minimal
loss of accuracy for orders of magnitude of
compression~\cite{han2015deep:binary}. This property allows these applications
to scale to datasets which might not otherwise fit in physical memory.  Common
to all of these approaches, however, is that the computation of distances
involves full-precision floating-point operations, often to the detriment of 
performance and memory overhead.

\begin{figure}[t]
  \centering
    \includegraphics[scale=0.48]{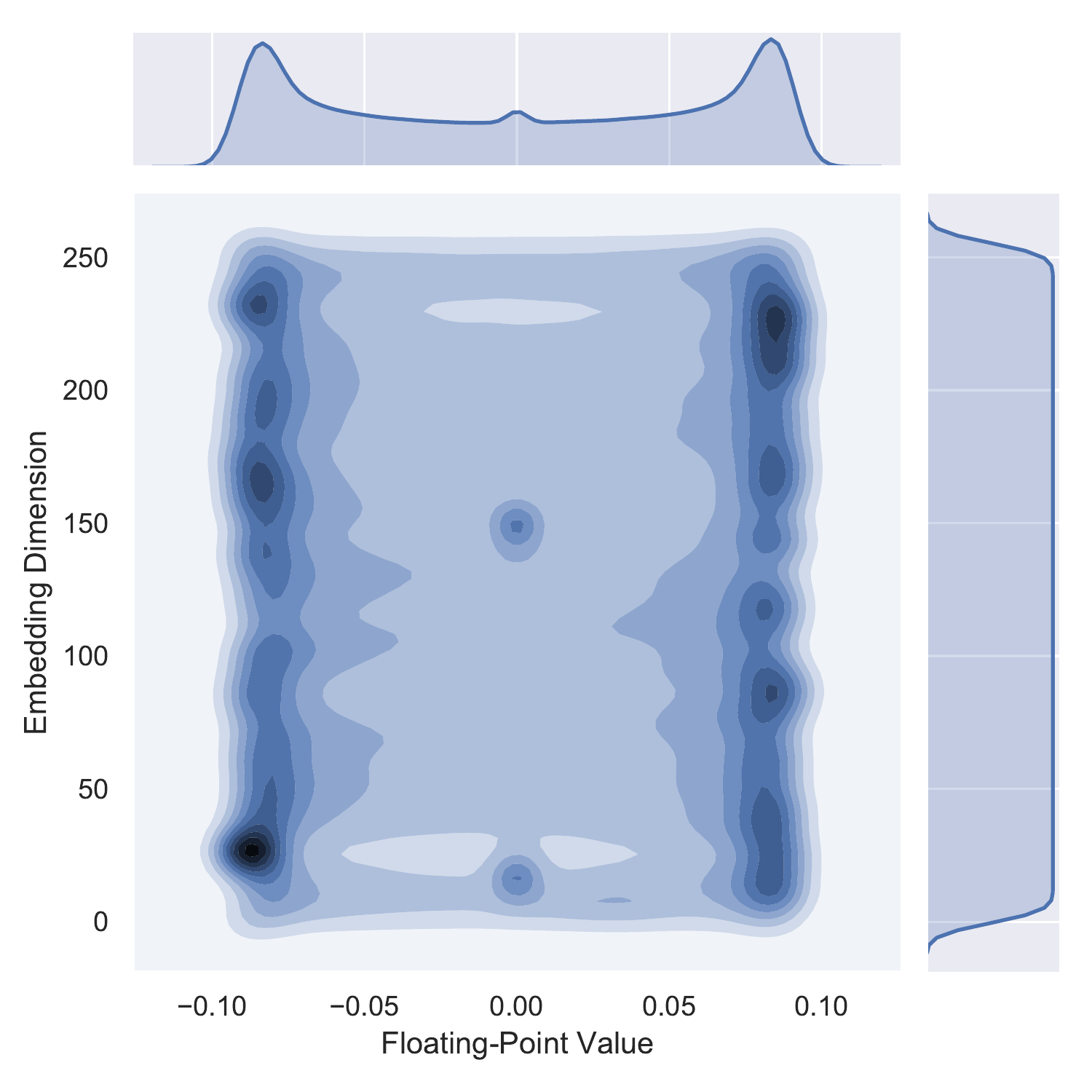}
    \caption{Distribution of values in a dataset of 100 million product embeddings derived from a large e-commerce catalog. We observe that the values comprising these embeddings cluster in a very narrow band.}
    \label{fig:dist}
\end{figure}

In practice, we observe that \knn{} corpora found in both
industry~\cite{li2017fexipro} and open source datasets~\cite{annbench} are dramatically
over-provisioned in terms of the range and precision that they can represent.
For example, Figure~\ref{fig:dist} plots the distribution of feature values
from a product embedding dataset used by a large e-commerce search engine. The data is
highly-structured, with values observed exclusively in the range $(-.125,
.125)$. The distribution is consistent across dimensions, with the majority of
values (50\%) observed in the narrow band $\pm(.125,.08)$. 

This observation suggests a novel vector quantization method which focuses on
feature-wise quantization. Given a corpus of vectors, we perform a data-driven
analysis to determine the range of values that appear in practice and then
quantize both the corpus and the distance function we intend to use into the
lowest-precision integer domain that can capture the relative positions of the
points in that set.  As long as the quantization is distance preserving in the
quantized space, the loss in recall is minimized.  As a simple example,
consider the following one-dimensional vectors (i.e.
points) $\{1.23,2.34,3.09,1.4e7\}$. We can safely map these points into a
smaller integer space of $\{1,2,3,4\}$ without considerable loss of recall in
nearest-neighbors. The third point remains the nearest neighbor of the fourth,
even after its value is altered by seven orders of magnitude.

This approach has two major consequences.  First, by using a compact
representation (say, {\tt int8}), it is possible to reduce memory overhead and
scale to larger datasets.  This has direct implications on storage-specific
design decisions and, hence, efficiency since an algorithm that stores vectors on
disk will present significantly different performance properties than one which
stores them in physical memory.  Secondly, primitive computations involving
integer data types can be more efficient than their floating-point
counterparts.  This implies that our approach can be combined 
with existing indexing-based KNN
frameworks~\cite{malkov2018efficient:hnsw,johnson2019billion:faiss} as a
mechanism for replacing full-precision vectors with compact integer
alternatives, and obtaining reductions in runtime overhead. 

\section{Related Work}
\label{sec:rel-work}


A large body of work in the \knn{} literature focuses on non-quantization-based approximate
methods for performing inner product and nearest neighbor search.  Locality
sensitive hashing~\cite{andoni2015practical,shrivastava2014asymmetric}, tree-based methods~\cite{muja2014scalable-tree, dasgupta2008random-tree}, and graph-based methods~\cite{harwood2016fanng-graph, malkov2018efficient:hnsw} focus on
achieving non-exhaustive search by partitioning the search space. 
%
Other work~\cite{he2013k:quantization} focuses on learning short binary codes
which can be searched using Hamming distances.  


The seminal work which describes the application of quantization to \knn{}
is product quantization~\cite{jegou2010product:pq}.  In
product quantization, the original vector space is decomposed into a Cartesian
product of lower dimensional subspaces, and vector quantization is performed in
each subspace independently.  Vector quantization approximates a
vector \(x \in \mathbb{R}^{d}\) by finding the closest quantizer in a codebook
\(\mathbf{C}\) 
$$
\phi_{V Q}(x ; \mathbf{C})=\underset{c \in\left\{\mathbf{C}_{j}\right\}}{\operatorname{argmin}}\|x-c\|_{2}
$$
where \(\mathbf{C} \in \mathbb{R}^{d \times m}\) is a vector quantization
codebook with \(m\) codewords, and the \(j\)-th column \(\mathbf{C}_{j}\)
represents the \(j\)-th quantizer.  This approach can be extended to \(K\)
subspaces.  
Numerous improvements have been proposed to this approach with the aim of
reducing subspace dependencies~\cite{ge2013optimized:opq,
norouzi2013cartesian}, including additive
quantization~\cite{babenko2014additive}, composite quantization
~\cite{zhang2014:composite,zhang2015sparse:composite} and stacked
quantization~\cite{martinez2014stacked}.  Our approach is complementary to
these approaches in the sense that one can either replace the original dataset
with low-precision quantized vectors or use it after the codebook mapping step for calculating the distance computations at query time.

A closely related recent work is that of \citet{guo2020accelerating}, who proposed an anisotropic vector quantization scheme called ScaNN which, using an insight similar to ours, modifies the product quantization objective function to minimize the relative distance between vectors as opposed to the reconstruction distance. We note that our work and ScaNN are complementary. We make the same observations, but pursue them differently. ScaNN reformulates existing quantization approaches in terms of a new, relative distance-preserving optimization criteria. In this work, we do not make modifications to the underlying algorithms; instead, we modify them at the implementation level by utilizing efficient integer computations.

\section{Low Precision Vector Quantization}
\label{sec:approach}

\subsection{Problem Formulation}

We propose a quantization family $(\mathcal{Q}, \phi)$ which is a combination
of quantization function $\mathcal{Q} : \mathbb{R}^d \rightarrow \mathbb{Z}^d$ and
distance function $\phi : \mathbb{Z}^d \times \mathbb{Z}^d \rightarrow
\mathbb{Z}$.  Following the notation in~\cite{bachrach2014speeding:xbox} we
first introduce the notion of a search problem.
\begin{definition}
  A \textbf{search problem} \(S(\I, \Q, d)\) consists of a set of \(n\) items
  \(\I=\left\{x_{1}, x_{2}, \ldots, x_{n}\right\},\) a set of queries
  \(\Q\) and a search function
$
d: \I \times \Q \rightarrow\{1,2, \ldots, n\}
$ such that the search function $d$ retrieves the index of an item in $\I$ for a given query $\q \in \Q$. 
\end{definition}

In this work, we focus on the maximum inner product (MIP) search problem, namely the task of retrieving the items having the largest inner product with a query vector $q \in \mathbb{R}^d$. With this definition in hand, we can formalize the concept of a quantization between search problems, which is a preprocessing step designed to improve search efficiency.


\begin{definition}
\label{def:quant}
  \(S_{2}\left(\mathcal{I}^{\prime}, \q^{\prime}, d_{2}\right)\) is a
  \textbf{quantization} of an original search problem
  \(S_{1}\left(\I, \q, d_{1}\right)\) if there exist functions \(g:\I \rightarrow
  \I^{\prime}\) and \(h: \q \rightarrow \q^{\prime}\) such that \(S_{2}\)
  is partial distance preserving meaning.
$$
\text{if } d_{1}(a,\q) < d_{1}(b,\q)  \text { then } d_{2}(g(a),h(\q)) \leq d_{2}(g(b),h(\q))
$$
\end{definition}

Distance preservation implies that if the triangle inequality
holds in the original distance space, it still holds in the quantized space.
Thus, metrics which satisfy this property remain valid for
quantized vectors. Using the standard definition of \textit{recall} (the
fraction of nearest neighbors retained by the quantized computation) we note
that the loss of recall for these metrics arises solely from the equality relaxation. We discuss the practical consequences of this property further in our experimental evaluations.

\subsection{Methodology}

For MIP, we propose a simple quantization function which ensures
that relative distances for nearest neighbors are preserved and accepts some
errors for instances which are further apart.  For a given bit-width $B$, we use a clamped
linear function with constants to quantize the $i$-th dimension of a vector $x$ as described below
\begin{equation}
    \mathcal{Q}(x^i) = 
    \begin{cases}
            \left\lfloor 2^B \cdot \frac{x^i-k^i}{S_e^i-S_b^i} \right \rfloor & \text{ if } x^i \in [S_b^i, S_e^i]\\
            -2^{B-1} & \text{ if } x^i < S_b^i \\
            2^{B-1} & \text{ if } x^i > S_e^i \\
    \end{cases}
\label{eqn:quantization}
\end{equation}


where $k^i$, $S_e^i$, $S_b^i$ are non-negative normalizing constants set per dimension. This non-negativity property ensures that distances are partially preserved as described in Definition \ref{def:quant}. Our goal is to learn $\mathcal{Q}(x^i)$. Towards this end, we assume that the elements of a vector are independent of one
another, and vectors are conditionally independent given an element. These are
strong assumptions, but as we show in Section~\ref{sec:evaluation}, are
borne out in practice.  This leads to the following expression for the
likelihood of the data set $I$.
$$
{\theta}=\underset{{\theta}}{\arg \max } \prod_{x \in I} \prod_{i} \mathbb{P}(x^i ; \theta)
$$

Parameterizing $\mathbb{P}$ as $\mathcal{N}(\mu, \sigma)$ we estimate
$\theta^i$ for each dimension in $I$.  The values of $(\mu^i, \sigma^i)$ can
then be used to set the constants in Equation~\ref{eqn:quantization}.  Given a
budget of $2^B$ where $B$ is the number of bits per dimension, we set $S_b^i =
\mu^i - \sigma^i$, $S_e^i = \mu^i + \sigma^i$, and $k^i = \mu^i$. Extending
this approach to datasets with high interdimensional variance or a significant
numbers of outliers, would be straightforward if tedious, and require
additional normalization constants for both scale and offset.  


\section{Implementation}
\label{sec:implementation}


This section describes our simplifying assumptions based on the highly structured
nature of the datasets we consider. As noted in Section~\ref{sec:introduction},
these properties are commonplace in both the literature and industry.

\subsection{Interdimensional Uniformity}

As suggested by Figure~\ref{fig:dist}, interdimensional variance tends to
decrease for datasets with a large number of dimensions. Furthermore, many
datasets are normalized to the unit ball during preprocessing. Thus, for these low-variance datasets, we assume a constant mean $\mu$ and standard deviation $\sigma$ across all dimensions. 




\subsection{Intradimensional Uniformity}

For datasets with significantly low intradimensional variance, we relax the
definition of $S_b^i$ and $S_e^i$ to allow for higher precision. Specifically,
for low-variance dimensions, even modestly sized values of $B$ are sufficient
for covering the range of values which we observe in practice. Thus rather than
clamp $\mathcal{Q}(x^i)$ to a single standard deviation, we use the absolute
maximum value observed in a given dimension and rely on standard techniques to
discard outliers.

\section{Evaluation}
\label{sec:eval}

\subsection{Algorithms and Datasets}



\begin{table}[!htbp]
\begin{tabular}{*5c}
\toprule
Config (EFC, M) &  \multicolumn{2}{c}{Build Time} & \multicolumn{2}{c}{Memory (GB)}\\
\midrule
		{}   & \textbf{fp32}   & \textbf{int8}    & \textbf{fp32}   & \textbf{int8}\\
		
		300, 32 & 1h 22 min & 44 min &	83.45  &   37.36\\
		300, 48 & 1h 20 min & 50 min &	91.11  &  45.03 \\
		400, 32 & 1h 38 min & 59 min & 	83.45  & 37.36\\
		400, 48 & 1h 51 min & 1h 6 min & 	91.11  & 45.03B\\
		600, 32 & 2h 32 min &1h 20 min &	83.45  & 37.36\\
		600, 48 & 2h 55 min & 1h 40 min &	91.11   &  45.03\\
		700, 32 & 3h 0 min &	1h 42 min &  83.45  & 37.36\\
		700, 48 & 3h 22 min &1h 58 min &	91.11& 45.03\\

\bottomrule
\end{tabular}
\caption{Build time and memory for HNSWlib indices with different parameters, applied to a 60MM dataset. 
         Quantization reduces memory overhead and improves build time.
 } 
\label{fig:buildmem}
\end{table}

\label{sec:evaluation}

\begin{figure*}[t]

    \begin{minipage}[b]{1\textwidth}
        \includegraphics[scale=0.315]{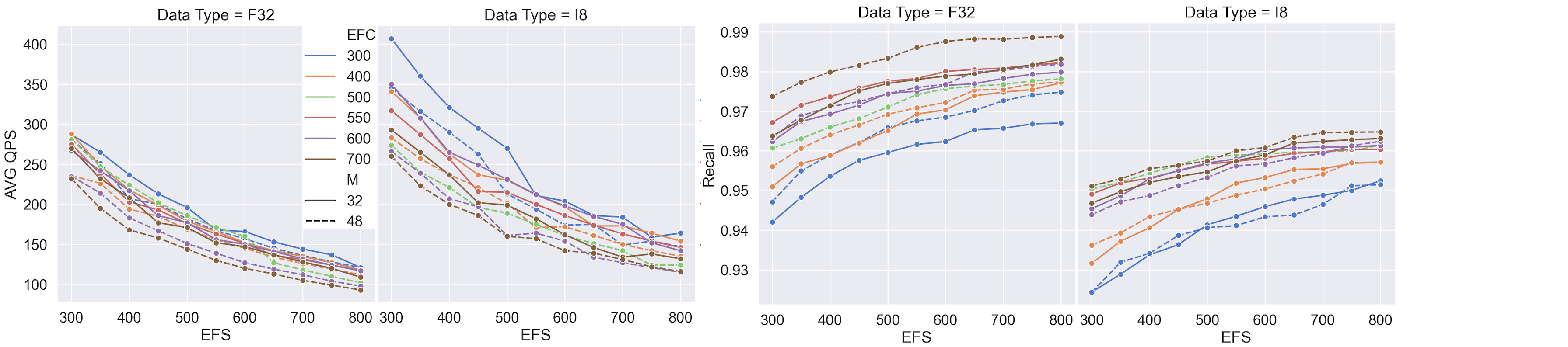}
            \caption{QPS and Recall versus EFS. Quantization results in higher overall QPS at only a slight decrease in recall.
            }
        \label{fig:recallqpsefs}
    \end{minipage}\hfill
\end{figure*}

We focus our evaluation on the Hierarchical Navigable Small World (HNSW) algorithm \cite{malkov2018efficient:hnsw}, which has consistently measured as one of the top-performing methods on public benchmarks and remains a popular choice for industry applications. HNSW is a graph traversal algorithm that involves constructing a layered graph from the set of vector embeddings. To measure the scalability of our proposed techniques on a real-world dataset, we focus our evaluation on a collection of 60 million vector representations of products sampled from the catalog of a large e-commerce search engine, using the approach described in \cite{nigam2019semantic} to learn the embeddings. We refer to this benchmark as PRODUCT60M. This benchmark also provides 1000 search queries represented by embeddings in the same semantic search space, which we used for recall measurements. In addition, to provide evidence of the generality of our approach, we also consider two additional popular \knn{} algorithms, namely FAISS
~\cite{JDH17} and NGT~\cite{DBLP:journals/corr/abs-1810-07355}, and two public benchmark datasets from ~\citet{DBLP:journals/corr/abs-1807-05614}. In all of our experiments, we fix $k$, the number of nearest neighbors to retrieve, to 100. 
We analyzed each of these algorithms with and without compressing the embedding
vectors and report results relating to (1) memory footprint, (2) indexing time,
(3) throughput, and (4) recall.  All experiments were run on a single AWS
r5n.24xlarge instance with Xeon(R) Platinum CPU 2.50GHz. For measuring indexing
time, we used all available CPU cores.  For measuring throughput, we used a
single thread. 

\subsection{Experimental Design}


We used an open source implementation of HNSW known as HNSWLib\footnote{https://github.com/nmslib/hnswlib}.  HNSWlib
supports only {\tt float32} for both indexing and the computation of inner
product distance. In order to use {\tt int8} for this analysis, we extended the
implementation to support both indexing and similarity search on dense
embedding vectors represented by {\tt int8} arrays.
%
HNSW uses three hyper parameters referred to as EFC, M, and EFS. These
parameters have an effect on memory, build time, recall, and throughput. The EFC parameter sets the number of closest elements to the query that are stored within each layer of the HNSW graph during construction time, implying that a higher EFC leads to a more accurate search index but at the cost of a longer build time; the M parameter determines the maximum number of outgoing edges in the graph and heavily influences the memory footprint of the search index; finally, the EFS parameter is analogous to EFC except that it controls the number of neighbors dynamically stored at search time. In this study, we considered a range of values for each hyper parameter when we
measured performance. We used two M values, 32 and 48, values from 300
to 700 in increments of 100 for EFC, and 300 to 800 in increments of 50 for
EFS. 

\subsection{Evaluation Metrics}
We divide our metrics into two distinct categories: search performance and
search quality. For performance, we focused on \emph{throughput}, the number of
queries the algorithm can process per unit of time, and index \emph{build
time}. We measure throughput by evaluating a test query, 
recording its execution time with no other processes running, and then returning the
reciprocal value as a measure of queries per second (QPS). Following the standard in the literature, our primary metric for measuring search quality is \emph{recall}, defined as $\frac{|S_E \cap
S_A|}{|S_E|}$ where $S_E$ is the set of results returned by an exact \knn{}
search and $S_A$ is the items retrieved by our approximation algorithm.

\subsection{Build time and Memory}
\label{eval:buildmem}
We built two groups of HNSW indexes for each of the hyper parameter
combinations reported above. The first group was built using the original
floating point embedding vectors. The second group was built by compressing
those vectors using the method described in Section~\ref{sec:implementation}.
Table~\ref{fig:buildmem} summarizes our observations on build time and memory
footprint for the two groups, and shows a noticeable improvement for the
compressed indexes.  Specifically, the {\tt int8} indices required
approximately half the memory and build time of the {\tt float32} indices. We
note that the transition from {\tt float32} to {\tt int8} did not produce a
linear decrease in memory. This is due to the overhead of HNSW's indexing graph
data structure which consists entirely of native pointers.  

\subsection{Throughput}
\label{eval:qps}

Figure~\ref{fig:recallqpsefs} plots QPS as a function of the HNSW search
parameter, EFS. The results were computed for several 
parameter settings which have a non-trivial effect on throughput.  The {\tt
int8} indexes produce higher throughput than the {\tt float32} indices. This phenomenon is likely due to the reduced overhead of native {\tt int8} operations versus
native {\tt float32} operations. We observed that there is a negative
non-linear association between the search parameter EFS and QPS. This is
expected, as higher EFS values correspond to longer search times.  

\subsection{Recall}
\label{eval:recall}
As noted in Section~\ref{sec:approach}, the primary source of recall loss for
distance metrics which obey the triangle inequality is equality relaxation.
Figure~\ref{fig:recallqpsefs} reports the relationship between HNSW
parameters and recall.  The original {\tt float32} indices achieve a higher
overall recall than the compressed {\tt int8} indices, but only by around 2\%.
Furthermore, as the value of the search parameter EFS increases, so too does
recall. This observation suggests that our quantization scheme achieves its
goal of preserving relative distances for nearest neighbors at the cost of
accepting some aliasing errors for points which are further apart. We also see similar small losses in recall when we evaluated on additional algorithms and datasets, as shown in Tables \ref{tbl:faiss} and \ref{tbl:ngt}. 

\subsection{Extended Evaluation}
Our primary evaluation used HNSWlib along with the PRODUCT60M dataset due to the
observation that HNSW offered the best recall-QPS tradeoff. However, to provide some evidence for the generality of our
approach and its expected performance on other domains and algorithms, we
consider two other KNN algorithms and public datasets, specifically the FAISS
~\cite{JDH17} exact search implementation and Neighborhood Graph and Tree (NGT)~\cite{DBLP:journals/corr/abs-1810-07355}. 

  \begin{table}[t]
  	\begin{tabular}{*4c}
  		\toprule
  		Dataset & Distance type &  \multicolumn{2}{c}{Recall@100} \\
  		\midrule
  		{}   &{}  & \textbf{fp32}   & \textbf{int8}\\
  		SIFT & L2 & 0.999 & 0.974\\
  		Glove100 & Angular & 1.0 & 0.943\\
  		PRODUCT60M & IP & 1.0 & 0.983\\
  		\bottomrule
  	\end{tabular}
          \caption{({\tt float32}) versus compressed ({\tt int8}) datasets using FAISS exhaustive nearest neighbor search.} 
  	\label{tbl:faiss}
  \end{table}

  \begin{table}[t]
	\begin{tabular}{*4c}
		\toprule
		Dataset & Distance type &  \multicolumn{2}{c}{Recall@100} \\
		\midrule
		{}   &{}  & \textbf{fp32}   & \textbf{int8}\\
		SIFT & L2 & 0.999 & 0.974\\
		Glove100 & Angular & 0.991 & 0.943\\
		PRODUCT60M & IP & 0.972 & 0.949\\
		\bottomrule
	\end{tabular}
          \caption{({\tt float32}) versus compressed ({\tt int8}) data sets using Neighborhood Graph and Tree (NGT).} 
	\label{tbl:ngt}
\end{table}

In addition to the search dataset described in Section~\ref{sec:implementation}
we used two public data sets from the KNN benchmark
~\cite{DBLP:journals/corr/abs-1807-05614}. The SIFT and Glove100 datasets have
dimensions of 128 and 100 respectively, and consist of approximately 1 million
vectors each. Each data set has its own 1000 vector test set along with 100
true neighbors for each test vector. 
Table~\ref{tbl:faiss} reports the performance of FAISS both before and after
quantization. Table ~\ref{tbl:ngt} reports the same for NGT. In both cases we
were able to achieve comparable improvements in runtime and memory consumption
at a slightly larger, though still modest, decrease in recall of $2-6\%$.

\section{Future Work and Conclusion}
\label{sec:conclusion}

We propose a low precision quantization technique based on the observation
that most real-world vector datasets are concentrated in a narrow range that
does not require full {\tt float32} precision. We applied this technique to the HNSW \knn{} algorithm and obtained a $60\%$ reduction in memory
overhead and a $30\%$ increase in throughput at only a $2\%$ reduction in
recall. We also provided evidence of the generality of our method by obtaining similar performance improvements on the FAISS exact search and NGT frameworks. For future work, we hope to further demonstrate the generality of our technique by evaluating on additional \knn{} algorithms and real-world benchmark datasets and extend our evaluation to additional distance metrics. 



\bibliography{reference}


\begin{thebibliography}{31}


\ifx \showCODEN    \undefined \def \showCODEN     #1{\unskip}     \fi
\ifx \showDOI      \undefined \def \showDOI       #1{#1}\fi
\ifx \showISBNx    \undefined \def \showISBNx     #1{\unskip}     \fi
\ifx \showISBNxiii \undefined \def \showISBNxiii  #1{\unskip}     \fi
\ifx \showISSN     \undefined \def \showISSN      #1{\unskip}     \fi
\ifx \showLCCN     \undefined \def \showLCCN      #1{\unskip}     \fi
\ifx \shownote     \undefined \def \shownote      #1{#1}          \fi
\ifx \showarticletitle \undefined \def \showarticletitle #1{#1}   \fi
\ifx \showURL      \undefined \def \showURL       {\relax}        \fi
\providecommand\bibfield[2]{#2}
\providecommand\bibinfo[2]{#2}
\providecommand\natexlab[1]{#1}
\providecommand\showeprint[2][]{arXiv:#2}

\bibitem[\protect\citeauthoryear{Amato and Falchi}{Amato and Falchi}{2010}]%
        {10.1145/1862344.1862360}
\bibfield{author}{\bibinfo{person}{Giuseppe Amato} {and}
  \bibinfo{person}{Fabrizio Falchi}.} \bibinfo{year}{2010}\natexlab{}.
\newblock \showarticletitle{KNN Based Image Classification Relying on Local
  Feature Similarity}. In \bibinfo{booktitle}{\emph{Proceedings of the Third
  International Conference on SImilarity Search and APplications}} (Istanbul,
  Turkey) \emph{(\bibinfo{series}{SISAP ’10})}.
  \bibinfo{publisher}{Association for Computing Machinery},
  \bibinfo{address}{New York, NY, USA}, \bibinfo{pages}{101–108}.
\newblock
\showISBNx{9781450304207}
\urldef\tempurl%
\url{https://doi.org/10.1145/1862344.1862360}
\showDOI{\tempurl}


\bibitem[\protect\citeauthoryear{Andoni, Indyk, Laarhoven, Razenshteyn, and
  Schmidt}{Andoni et~al\mbox{.}}{2015}]%
        {andoni2015practical}
\bibfield{author}{\bibinfo{person}{Alexandr Andoni}, \bibinfo{person}{Piotr
  Indyk}, \bibinfo{person}{Thijs Laarhoven}, \bibinfo{person}{Ilya
  Razenshteyn}, {and} \bibinfo{person}{Ludwig Schmidt}.}
  \bibinfo{year}{2015}\natexlab{}.
\newblock \showarticletitle{Practical and optimal LSH for angular distance}. In
  \bibinfo{booktitle}{\emph{Advances in neural information processing
  systems}}. \bibinfo{pages}{1225--1233}.
\newblock


\bibitem[\protect\citeauthoryear{Aum{\"{u}}ller, Bernhardsson, and
  Faithfull}{Aum{\"{u}}ller et~al\mbox{.}}{2018}]%
        {DBLP:journals/corr/abs-1807-05614}
\bibfield{author}{\bibinfo{person}{Martin Aum{\"{u}}ller},
  \bibinfo{person}{Erik Bernhardsson}, {and} \bibinfo{person}{Alexander~John
  Faithfull}.} \bibinfo{year}{2018}\natexlab{}.
\newblock \showarticletitle{ANN-Benchmarks: {A} Benchmarking Tool for
  Approximate Nearest Neighbor Algorithms}.
\newblock \bibinfo{journal}{\emph{CoRR}}  \bibinfo{volume}{abs/1807.05614}
  (\bibinfo{year}{2018}).
\newblock
\showeprint[arxiv]{1807.05614}
\urldef\tempurl%
\url{http://arxiv.org/abs/1807.05614}
\showURL{%
\tempurl}


\bibitem[\protect\citeauthoryear{Aum{\"{u}}ller, Bernhardsson, and
  Faithfull}{Aum{\"{u}}ller et~al\mbox{.}}{2020}]%
        {annbench}
\bibfield{author}{\bibinfo{person}{Martin Aum{\"{u}}ller},
  \bibinfo{person}{Erik Bernhardsson}, {and} \bibinfo{person}{Alexander~John
  Faithfull}.} \bibinfo{year}{2020}\natexlab{}.
\newblock \showarticletitle{ANN-Benchmarks: {A} benchmarking tool for
  approximate nearest neighbor algorithms}.
\newblock \bibinfo{journal}{\emph{Inf. Syst.}}  \bibinfo{volume}{87}
  (\bibinfo{year}{2020}).
\newblock
\urldef\tempurl%
\url{https://doi.org/10.1016/j.is.2019.02.006}
\showDOI{\tempurl}


\bibitem[\protect\citeauthoryear{Babenko and Lempitsky}{Babenko and
  Lempitsky}{2014}]%
        {babenko2014additive}
\bibfield{author}{\bibinfo{person}{Artem Babenko} {and} \bibinfo{person}{Victor
  Lempitsky}.} \bibinfo{year}{2014}\natexlab{}.
\newblock \showarticletitle{Additive quantization for extreme vector
  compression}. In \bibinfo{booktitle}{\emph{Proceedings of the IEEE Conference
  on Computer Vision and Pattern Recognition}}. \bibinfo{pages}{931--938}.
\newblock


\bibitem[\protect\citeauthoryear{Bachrach, Finkelstein, Gilad-Bachrach, Katzir,
  Koenigstein, Nice, and Paquet}{Bachrach et~al\mbox{.}}{2014}]%
        {bachrach2014speeding:xbox}
\bibfield{author}{\bibinfo{person}{Yoram Bachrach}, \bibinfo{person}{Yehuda
  Finkelstein}, \bibinfo{person}{Ran Gilad-Bachrach}, \bibinfo{person}{Liran
  Katzir}, \bibinfo{person}{Noam Koenigstein}, \bibinfo{person}{Nir Nice},
  {and} \bibinfo{person}{Ulrich Paquet}.} \bibinfo{year}{2014}\natexlab{}.
\newblock \showarticletitle{Speeding up the xbox recommender system using a
  euclidean transformation for inner-product spaces}. In
  \bibinfo{booktitle}{\emph{Proceedings of the 8th ACM Conference on
  Recommender systems}}. \bibinfo{pages}{257--264}.
\newblock


\bibitem[\protect\citeauthoryear{Dasgupta and Freund}{Dasgupta and
  Freund}{2008}]%
        {dasgupta2008random-tree}
\bibfield{author}{\bibinfo{person}{Sanjoy Dasgupta} {and} \bibinfo{person}{Yoav
  Freund}.} \bibinfo{year}{2008}\natexlab{}.
\newblock \showarticletitle{Random projection trees and low dimensional
  manifolds}. In \bibinfo{booktitle}{\emph{Proceedings of the fortieth annual
  ACM symposium on Theory of computing}}. \bibinfo{pages}{537--546}.
\newblock


\bibitem[\protect\citeauthoryear{Garcia, Debreuve, and Barlaud}{Garcia
  et~al\mbox{.}}{2008}]%
        {garcia2008fast:knn-gpu}
\bibfield{author}{\bibinfo{person}{Vincent Garcia}, \bibinfo{person}{Eric
  Debreuve}, {and} \bibinfo{person}{Michel Barlaud}.}
  \bibinfo{year}{2008}\natexlab{}.
\newblock \showarticletitle{Fast k nearest neighbor search using GPU}. In
  \bibinfo{booktitle}{\emph{2008 IEEE Computer Society Conference on Computer
  Vision and Pattern Recognition Workshops}}. IEEE, \bibinfo{pages}{1--6}.
\newblock


\bibitem[\protect\citeauthoryear{Ge, He, Ke, and Sun}{Ge et~al\mbox{.}}{2013}]%
        {ge2013optimized:opq}
\bibfield{author}{\bibinfo{person}{Tiezheng Ge}, \bibinfo{person}{Kaiming He},
  \bibinfo{person}{Qifa Ke}, {and} \bibinfo{person}{Jian Sun}.}
  \bibinfo{year}{2013}\natexlab{}.
\newblock \showarticletitle{Optimized product quantization for approximate
  nearest neighbor search}. In \bibinfo{booktitle}{\emph{Proceedings of the
  IEEE Conference on Computer Vision and Pattern Recognition}}.
  \bibinfo{pages}{2946--2953}.
\newblock


\bibitem[\protect\citeauthoryear{Guo, Sun, Lindgren, Geng, Simcha, Chern, and
  Kumar}{Guo et~al\mbox{.}}{2020}]%
        {guo2020accelerating}
\bibfield{author}{\bibinfo{person}{Ruiqi Guo}, \bibinfo{person}{Philip Sun},
  \bibinfo{person}{Erik Lindgren}, \bibinfo{person}{Quan Geng},
  \bibinfo{person}{David Simcha}, \bibinfo{person}{Felix Chern}, {and}
  \bibinfo{person}{Sanjiv Kumar}.} \bibinfo{year}{2020}\natexlab{}.
\newblock \showarticletitle{Accelerating large-scale inference with anisotropic
  vector quantization}. In \bibinfo{booktitle}{\emph{International Conference
  on Machine Learning}}. PMLR, \bibinfo{pages}{3887--3896}.
\newblock


\bibitem[\protect\citeauthoryear{Han, Mao, and Dally}{Han
  et~al\mbox{.}}{2015}]%
        {han2015deep:binary}
\bibfield{author}{\bibinfo{person}{Song Han}, \bibinfo{person}{Huizi Mao},
  {and} \bibinfo{person}{William~J Dally}.} \bibinfo{year}{2015}\natexlab{}.
\newblock \showarticletitle{Deep compression: Compressing deep neural networks
  with pruning, trained quantization and huffman coding}.
\newblock \bibinfo{journal}{\emph{arXiv preprint arXiv:1510.00149}}
  (\bibinfo{year}{2015}).
\newblock


\bibitem[\protect\citeauthoryear{Harwood and Drummond}{Harwood and
  Drummond}{2016}]%
        {harwood2016fanng-graph}
\bibfield{author}{\bibinfo{person}{Ben Harwood} {and} \bibinfo{person}{Tom
  Drummond}.} \bibinfo{year}{2016}\natexlab{}.
\newblock \showarticletitle{Fanng: Fast approximate nearest neighbour graphs}.
  In \bibinfo{booktitle}{\emph{Proceedings of the IEEE Conference on Computer
  Vision and Pattern Recognition}}. \bibinfo{pages}{5713--5722}.
\newblock


\bibitem[\protect\citeauthoryear{He, Wen, and Sun}{He et~al\mbox{.}}{2013}]%
        {he2013k:quantization}
\bibfield{author}{\bibinfo{person}{Kaiming He}, \bibinfo{person}{Fang Wen},
  {and} \bibinfo{person}{Jian Sun}.} \bibinfo{year}{2013}\natexlab{}.
\newblock \showarticletitle{K-means hashing: An affinity-preserving
  quantization method for learning binary compact codes}. In
  \bibinfo{booktitle}{\emph{Proceedings of the IEEE conference on computer
  vision and pattern recognition}}. \bibinfo{pages}{2938--2945}.
\newblock


\bibitem[\protect\citeauthoryear{Huang, He, Gao, Deng, Acero, and Heck}{Huang
  et~al\mbox{.}}{2013}]%
        {huang2013learning}
\bibfield{author}{\bibinfo{person}{Po-Sen Huang}, \bibinfo{person}{Xiaodong
  He}, \bibinfo{person}{Jianfeng Gao}, \bibinfo{person}{Li Deng},
  \bibinfo{person}{Alex Acero}, {and} \bibinfo{person}{Larry Heck}.}
  \bibinfo{year}{2013}\natexlab{}.
\newblock \showarticletitle{Learning Deep Structured Semantic Models for Web
  Search using Clickthrough Data}. \bibinfo{publisher}{ACM International
  Conference on Information and Knowledge Management (CIKM)}.
\newblock
\urldef\tempurl%
\url{https://www.microsoft.com/en-us/research/publication/learning-deep-structured-semantic-models-for-web-search-using-clickthrough-data/}
\showURL{%
\tempurl}


\bibitem[\protect\citeauthoryear{Iwasaki and Miyazaki}{Iwasaki and
  Miyazaki}{2018}]%
        {DBLP:journals/corr/abs-1810-07355}
\bibfield{author}{\bibinfo{person}{Masajiro Iwasaki} {and}
  \bibinfo{person}{Daisuke Miyazaki}.} \bibinfo{year}{2018}\natexlab{}.
\newblock \showarticletitle{Optimization of Indexing Based on k-Nearest
  Neighbor Graph for Proximity Search in High-dimensional Data}.
\newblock \bibinfo{journal}{\emph{CoRR}}  \bibinfo{volume}{abs/1810.07355}
  (\bibinfo{year}{2018}).
\newblock
\showeprint[arxiv]{1810.07355}
\urldef\tempurl%
\url{http://arxiv.org/abs/1810.07355}
\showURL{%
\tempurl}


\bibitem[\protect\citeauthoryear{Jegou, Douze, and Schmid}{Jegou
  et~al\mbox{.}}{2010}]%
        {jegou2010product:pq}
\bibfield{author}{\bibinfo{person}{Herve Jegou}, \bibinfo{person}{Matthijs
  Douze}, {and} \bibinfo{person}{Cordelia Schmid}.}
  \bibinfo{year}{2010}\natexlab{}.
\newblock \showarticletitle{Product quantization for nearest neighbor search}.
\newblock \bibinfo{journal}{\emph{IEEE transactions on pattern analysis and
  machine intelligence}} \bibinfo{volume}{33}, \bibinfo{number}{1}
  (\bibinfo{year}{2010}), \bibinfo{pages}{117--128}.
\newblock


\bibitem[\protect\citeauthoryear{Johnson, Douze, and J{\'e}gou}{Johnson
  et~al\mbox{.}}{2017}]%
        {JDH17}
\bibfield{author}{\bibinfo{person}{Jeff Johnson}, \bibinfo{person}{Matthijs
  Douze}, {and} \bibinfo{person}{Herv{\'e} J{\'e}gou}.}
  \bibinfo{year}{2017}\natexlab{}.
\newblock \showarticletitle{Billion-scale similarity search with GPUs}.
\newblock \bibinfo{journal}{\emph{arXiv preprint arXiv:1702.08734}}
  (\bibinfo{year}{2017}).
\newblock


\bibitem[\protect\citeauthoryear{Johnson, Douze, and J{\'e}gou}{Johnson
  et~al\mbox{.}}{2019}]%
        {johnson2019billion:faiss}
\bibfield{author}{\bibinfo{person}{Jeff Johnson}, \bibinfo{person}{Matthijs
  Douze}, {and} \bibinfo{person}{Herv{\'e} J{\'e}gou}.}
  \bibinfo{year}{2019}\natexlab{}.
\newblock \showarticletitle{Billion-scale similarity search with GPUs}.
\newblock \bibinfo{journal}{\emph{IEEE Transactions on Big Data}}
  (\bibinfo{year}{2019}).
\newblock


\bibitem[\protect\citeauthoryear{{Kim}, {Park}, {Ahn}, {Ko}, {Park}, and
  {Gallagher}}{{Kim} et~al\mbox{.}}{2017}]%
        {7894058}
\bibfield{author}{\bibinfo{person}{J. {Kim}}, \bibinfo{person}{C. {Park}},
  \bibinfo{person}{J. {Ahn}}, \bibinfo{person}{Y. {Ko}}, \bibinfo{person}{J.
  {Park}}, {and} \bibinfo{person}{J.~C. {Gallagher}}.}
  \bibinfo{year}{2017}\natexlab{}.
\newblock \showarticletitle{Real-time UAV sound detection and analysis system}.
  In \bibinfo{booktitle}{\emph{2017 IEEE Sensors Applications Symposium
  (SAS)}}. \bibinfo{pages}{1--5}.
\newblock


\bibitem[\protect\citeauthoryear{Li, Chan, Yiu, and Mamoulis}{Li
  et~al\mbox{.}}{2017}]%
        {li2017fexipro}
\bibfield{author}{\bibinfo{person}{Hui Li}, \bibinfo{person}{Tsz~Nam Chan},
  \bibinfo{person}{Man~Lung Yiu}, {and} \bibinfo{person}{Nikos Mamoulis}.}
  \bibinfo{year}{2017}\natexlab{}.
\newblock \showarticletitle{FEXIPRO: fast and exact inner product retrieval in
  recommender systems}. In \bibinfo{booktitle}{\emph{Proceedings of the 2017
  ACM International Conference on Management of Data}}.
  \bibinfo{pages}{835--850}.
\newblock


\bibitem[\protect\citeauthoryear{Malkov and Yashunin}{Malkov and
  Yashunin}{2018}]%
        {malkov2018efficient:hnsw}
\bibfield{author}{\bibinfo{person}{Yury~A Malkov} {and}
  \bibinfo{person}{Dmitry~A Yashunin}.} \bibinfo{year}{2018}\natexlab{}.
\newblock \showarticletitle{Efficient and robust approximate nearest neighbor
  search using hierarchical navigable small world graphs}.
\newblock \bibinfo{journal}{\emph{IEEE transactions on pattern analysis and
  machine intelligence}} (\bibinfo{year}{2018}).
\newblock


\bibitem[\protect\citeauthoryear{Martinez, Hoos, and Little}{Martinez
  et~al\mbox{.}}{2014}]%
        {martinez2014stacked}
\bibfield{author}{\bibinfo{person}{Julieta Martinez}, \bibinfo{person}{Holger~H
  Hoos}, {and} \bibinfo{person}{James~J Little}.}
  \bibinfo{year}{2014}\natexlab{}.
\newblock \showarticletitle{Stacked quantizers for compositional vector
  compression}.
\newblock \bibinfo{journal}{\emph{arXiv preprint arXiv:1411.2173}}
  (\bibinfo{year}{2014}).
\newblock


\bibitem[\protect\citeauthoryear{Muja and Lowe}{Muja and Lowe}{2014}]%
        {muja2014scalable-tree}
\bibfield{author}{\bibinfo{person}{Marius Muja} {and} \bibinfo{person}{David~G
  Lowe}.} \bibinfo{year}{2014}\natexlab{}.
\newblock \showarticletitle{Scalable nearest neighbor algorithms for high
  dimensional data}.
\newblock \bibinfo{journal}{\emph{IEEE transactions on pattern analysis and
  machine intelligence}} \bibinfo{volume}{36}, \bibinfo{number}{11}
  (\bibinfo{year}{2014}), \bibinfo{pages}{2227--2240}.
\newblock


\bibitem[\protect\citeauthoryear{Nigam, Song, Mohan, Lakshman, Ding, Shingavi,
  Teo, Gu, and Yin}{Nigam et~al\mbox{.}}{2019}]%
        {nigam2019semantic}
\bibfield{author}{\bibinfo{person}{Priyanka Nigam}, \bibinfo{person}{Yiwei
  Song}, \bibinfo{person}{Vijai Mohan}, \bibinfo{person}{Vihan Lakshman},
  \bibinfo{person}{Weitian Ding}, \bibinfo{person}{Ankit Shingavi},
  \bibinfo{person}{Choon~Hui Teo}, \bibinfo{person}{Hao Gu}, {and}
  \bibinfo{person}{Bing Yin}.} \bibinfo{year}{2019}\natexlab{}.
\newblock \showarticletitle{Semantic product search}. In
  \bibinfo{booktitle}{\emph{Proceedings of the 25th ACM SIGKDD International
  Conference on Knowledge Discovery \& Data Mining}}.
  \bibinfo{pages}{2876--2885}.
\newblock


\bibitem[\protect\citeauthoryear{Norouzi and Fleet}{Norouzi and Fleet}{2013}]%
        {norouzi2013cartesian}
\bibfield{author}{\bibinfo{person}{Mohammad Norouzi} {and}
  \bibinfo{person}{David~J Fleet}.} \bibinfo{year}{2013}\natexlab{}.
\newblock \showarticletitle{Cartesian k-means}. In
  \bibinfo{booktitle}{\emph{Proceedings of the IEEE Conference on computer
  Vision and Pattern Recognition}}. \bibinfo{pages}{3017--3024}.
\newblock


\bibitem[\protect\citeauthoryear{Shrivastava and Li}{Shrivastava and
  Li}{2014}]%
        {shrivastava2014asymmetric}
\bibfield{author}{\bibinfo{person}{Anshumali Shrivastava} {and}
  \bibinfo{person}{Ping Li}.} \bibinfo{year}{2014}\natexlab{}.
\newblock \showarticletitle{Asymmetric LSH (ALSH) for sublinear time maximum
  inner product search (MIPS)}. In \bibinfo{booktitle}{\emph{Advances in Neural
  Information Processing Systems}}. \bibinfo{pages}{2321--2329}.
\newblock


\bibitem[\protect\citeauthoryear{Suchal and N{\'a}vrat}{Suchal and
  N{\'a}vrat}{2010}]%
        {10.1007/978-3-642-15286-3_16}
\bibfield{author}{\bibinfo{person}{J{\'a}n Suchal} {and} \bibinfo{person}{Pavol
  N{\'a}vrat}.} \bibinfo{year}{2010}\natexlab{}.
\newblock \showarticletitle{Full Text Search Engine as Scalable k-Nearest
  Neighbor Recommendation System}. In \bibinfo{booktitle}{\emph{Artificial
  Intelligence in Theory and Practice III}},
  \bibfield{editor}{\bibinfo{person}{Max Bramer}} (Ed.).
  \bibinfo{publisher}{Springer Berlin Heidelberg}, \bibinfo{address}{Berlin,
  Heidelberg}, \bibinfo{pages}{165--173}.
\newblock
\showISBNx{978-3-642-15286-3}


\bibitem[\protect\citeauthoryear{Teflioudi and Gemulla}{Teflioudi and
  Gemulla}{2016}]%
        {teflioudi2016exact:lemp}
\bibfield{author}{\bibinfo{person}{Christina Teflioudi} {and}
  \bibinfo{person}{Rainer Gemulla}.} \bibinfo{year}{2016}\natexlab{}.
\newblock \showarticletitle{Exact and approximate maximum inner product search
  with lemp}.
\newblock \bibinfo{journal}{\emph{ACM Transactions on Database Systems (TODS)}}
  \bibinfo{volume}{42}, \bibinfo{number}{1} (\bibinfo{year}{2016}),
  \bibinfo{pages}{1--49}.
\newblock


\bibitem[\protect\citeauthoryear{{Wang}, {Liao}, and {Zhang}}{{Wang}
  et~al\mbox{.}}{2013}]%
        {6642782}
\bibfield{author}{\bibinfo{person}{B. {Wang}}, \bibinfo{person}{Q. {Liao}},
  {and} \bibinfo{person}{C. {Zhang}}.} \bibinfo{year}{2013}\natexlab{}.
\newblock \showarticletitle{Weight Based KNN Recommender System}. In
  \bibinfo{booktitle}{\emph{2013 5th International Conference on Intelligent
  Human-Machine Systems and Cybernetics}}, Vol.~\bibinfo{volume}{2}.
  \bibinfo{pages}{449--452}.
\newblock


\bibitem[\protect\citeauthoryear{Zhang, Du, and Wang}{Zhang
  et~al\mbox{.}}{2014}]%
        {zhang2014:composite}
\bibfield{author}{\bibinfo{person}{Ting Zhang}, \bibinfo{person}{Chao Du},
  {and} \bibinfo{person}{Jingdong Wang}.} \bibinfo{year}{2014}\natexlab{}.
\newblock \showarticletitle{Composite Quantization for Approximate Nearest
  Neighbor Search.}. In \bibinfo{booktitle}{\emph{ICML}},
  Vol.~\bibinfo{volume}{2}. \bibinfo{pages}{3}.
\newblock


\bibitem[\protect\citeauthoryear{Zhang, Qi, Tang, and Wang}{Zhang
  et~al\mbox{.}}{2015}]%
        {zhang2015sparse:composite}
\bibfield{author}{\bibinfo{person}{Ting Zhang}, \bibinfo{person}{Guo-Jun Qi},
  \bibinfo{person}{Jinhui Tang}, {and} \bibinfo{person}{Jingdong Wang}.}
  \bibinfo{year}{2015}\natexlab{}.
\newblock \showarticletitle{Sparse composite quantization}. In
  \bibinfo{booktitle}{\emph{Proceedings of the IEEE Conference on Computer
  Vision and Pattern Recognition}}. \bibinfo{pages}{4548--4556}.
\newblock


\end{thebibliography}


\end{document}